# Comparative Analysis of Widely use Object-Oriented Languages

Muhammad Shoaib Farooq[1], Taymour zaman Khan[1]
[1]Department of Artificial Intelligence, School of System and Technology, University of Management and Technology, Lahore, 54000, Pakistan
Corresponding author: Muhammad Shoaib Farooq (shoaib.farooq@umt.edu.pk)\

**ABSTRACT** programming is an integral part of computer science discipline. Every day the programming environment is not only rapidly growing but also changing and languages are constantly evolving. Learning of object-oriented paradigm is compulsory in every computer science major so the choice of language to teach object-oriented principles is very important. Due to large pool of object-oriented languages, it is difficult to choose which should be the first programming language in order to teach object-oriented principles. Many studies shown which should be the first language to tech object-oriented concepts but there is no method to compare and evaluate these languages. In this article we proposed a comprehensive framework to evaluate the widely used object-oriented languages. The languages are evaluated basis of their technical and environmental features. Furthermore, we have constructed a scoring function based on proposed evaluation framework which provides us a language's quantitative score allow us to determine which language is acceptable as first object-oriented language to teach. Moreover, we have also calculated the conformance of widely used object-oriented languages.

## I. INTRODUCTION

Computer Programming is the core of computer science major degree. Computer programming follows several paradigms out of those paradigms the most widely used and most flexible is object-oriented paradigm**[1]**. From two decades object-oriented paradigm is used for real time and complex applications Object-oriented paradigm is widely used in every type of software. Object oriented paradigm provides the features like code reusability and reliability by providing strong features like encapsulation and abstraction and inheritance**[2]**. Object oriented paradigm provides ease of maintainability of application and rapid and robust development. Every high-level language supports object-oriented paradigm. Moreover, object oriented paradigm provides solution of the problem based on real word objects **[3]**. As the popularity of paradigm grows in the software industry the academics include the object-oriented programming course in their curriculum.

For the past decade many languages have been used to teach the concepts of object-oriented programming. Different languages are used by educational institutes to teach object-oriented programming. But most popular used languages used by educational institutes are Smalltalk,Eiffel,Java, Python#, C++ and according to research **[2]** some languages like Blue design specifically to teach object oriented programming. The study**[4]** shows the languages which design specifically for educational purposes are not clearly meet the industrial demand and not recommended by professors. Teaching object-oriented programming is difficult for teachers because it contains both theoretical and practical concepts and students needs to grasp both theoretical and practical aspects**[5]**. Because in most

Institutes the educators teach procedural paradigm programming and then shift students to object oriented programming. This paradigm shift not only causes problem but students also needs to learn the grammatical rules of new language**[6]**. Due these variety of opinions there needs a comprehensive and general framework for the evaluation of appropriate object-oriented languages. Many researchers tried to define criteria **[4], [7] [10]** but they more related to specific object oriented language and as per our information they are not very extensive.

The major focus of this study is to provide the comprehensive model to evaluate object-oriented languages on the basis of their object-oriented concepts. The framework is composed of language technical and non-technical features. The novelty of our work is while defining our evaluation framework we object oriented principle rather than taking language related concepts and evaluate each non-technical feature as well. After that we define a universal customizable scoring function and can be changed according to users' preference. The next contribution of our work is we ranked and evaluate currently most popular and widely used object-oriented languages using our proposed framework. The selection of these languages is based on the **[11]** rankings which can be seen in **Table 1**.The remaining paper is arranged as follows: after the introduction section we highlighted the related work. The proposed framework discussed in detail under the section "Proposed Framework and Analysis" in which we define each object-oriented feature and its sub feature and then evaluate each feature and its sub feature and after evaluation we define a scoring function. Finally, we present the conclusion and future direction of this research.

*Table 1 Tiobe Index Language Rankings*

| Language | Rating | Change |
|---|---|---|
| **Python** | 13.44% | +2.48% |
| **Java** | 11.59% | 0.40% |
| **C++** | 10.00% | +1.98% |
| **C#** | 5.65% | +0.82% |

## II. RELATED WORK

Many assessments of programming languages are published till today but mostly efforts were made on imperative and procedural languages. There have been few studies on object-oriented programming which discussed the criteria of learning object-oriented programming as well as different techniques but they are not discussed all the details. The research on object-oriented programming covered either a qualitative analysis or quantitative analysis.

Sandhu [12] tried to evaluate and compare five object oriented languages C++ ,Java,Python,Ruby and C#. He applies each object-oriented concept on language and compare the execution of time and memory used by code in bytes of each language. The research is helpful when selecting the project which we have critical response time and memory.

The research does not evaluate the languages on the basis of their object-oriented features and one language have better implementation of object-oriented principles but rather provide the memory and running time analysis.

In study **[8]** the research to show that eighter Java or Python is more suitable for beginners they performed the comparative analysis between Java and Python on the basis of syntax and features. The research also used the applications and describe advantages and disadvantages. The research shows the results by using quick sort and tic tac toe game algorithms and compare their run time.

The research discusses the object-oriented features of both languages but does not provide a in depth details and a comprehensive way to evaluate them. The study is limited to only these two languages which maps on every object-oriented language.

The study **[4]** provides the extensive evaluation criteria based on technical and non-technical features and also provide the framework on widely used first programming languages and also define a customizable scoring function but study mainly focus on imperative and procedural languages and lack in provide the extensive details of object oriented languages and their features and does not provide a in dept details about object oriented paradigm.

There has been research on object-oriented programming but research does not compare the programming languages on the basis of their object-oriented implementation and nature of languages. In [7] compare the four object-oriented languages C++ Oberon-2, Modula -3, Self but the problem with comparison is its too old and don't compare the languages on the basis of object-oriented implementation.

The study compares languages on non-technical features like syntax run time and which language supports big projects. The goal of paper is more about encourage and facilitate the programmers to learn the new object-oriented language if they have some previous experience on object-oriented language but there is not details about the principles of object oriented paradigm.

The study [13] also study the six object-oriented languages but the main objective of authors is comparing the language on the basis of code reusability memory consumption and efficiency and expressiveness and do not cover the again basic pillars of object-oriented programming our research compares technical and non-technical feature of these four popular and modern languages.

All the above discussed researches show the efforts to compare the object-oriented programming languages but there are no global criteria in which one can compare object-oriented languages. This gap provides the opportunity to provide the extensive method to compare and evaluate the object-oriented languages. In this research we try to define the framework along with scoring function to evaluate and compare each object-oriented language. The novelty of our is we not only define the criteria with help of available literature to evaluate each language but also define each object-oriented principle and its sub features in more general way to make this a more generalized which helped to compare the object-oriented languages.

## III. Proposed Evaluation Framework and Analysis

In this research our proposed evaluation criteria are based on two features technical and non-technical. In technical features we discuss the object-oriented concepts and their sb features which is shown in Figure 1 and in non-technical features we discuss environmental factors and readability according to object-oriented programming then we discuss their sub features. Firstly, we define the feature then discuss how language implemented a particular feature and how its sub feature formed by using the language and then evaluate it. Then rate these features and sub feature. We use four qualitative values Fully, Mostly, Partially and No. If a language full fills all the major requirements mostly indicate that some requirements are not met but most of them met partially means that vast portion of requirements are not met and only a few requirements met and No means not full fill at all the above qualitative measures has already been discussed in the study [14]. In order to define a feature and its sub feature and evaluate them with help of previous literature resources and contextual references and after all feature comparison assign each feature a weight and after classification and we calculate which language supports most object-oriented concepts by defining a proper scoring function.

## A) TECHNICAL FEATURES

In this section we discuss each technical feature in detail and define its sub feature and rate each feature and sub feature in our framework.

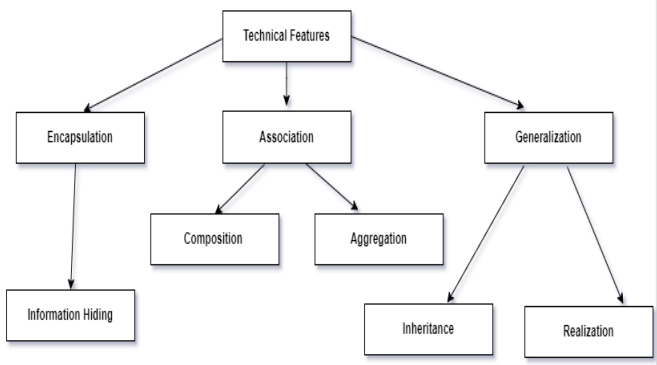

*Figure 1 Technical Features*

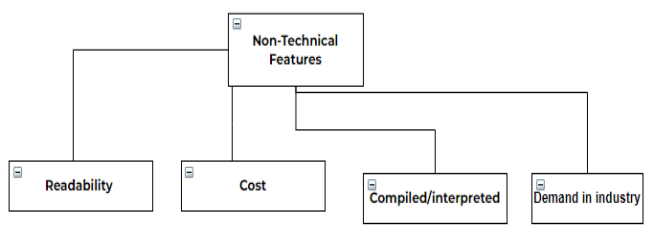

*Figure 2 Enviromental Features*

### 1) ENCAPSULATION

In object-oriented paradigm the notion encapsulation describes the data attributes and actions or methods which operates on data is combined in a single unit which is known as class and data is not accessible for outside world which data attributes should be private and in order to access data client must use public interface in the form of public functions. Encapsulation ensures the security of data and it provides the abstraction of complex implementation. The important benefit of using encapsulation we don't have to change client side if we change in business logic of public interfaces. In [15] book he defines the two types of encapsulations the first he abstracts datatypes encapsulations and the second is naming encapsulations. Abstract datatypes encapsulation further divides into single assembly and multiple assembly which is also known as naming encapsulation. We use these two types of encapsulations to evaluate the implementation of encapsulation offered by these four languages.

### a) ENCAPSULATION IN ABSTRACT DATA TYPES

This section explain encapsulation in abstract data types in selected four languages respectively. We will discuss the levels of encapsulation provide by each language and evaluate them on the basis of following parameters (I) implicit access modifier (II) properties (III) auto implement properties (IV) nested classes.

**Implicit access modifier :** The first thing we evaluate is what default access specifier is provided by the language to its attributes because it enforce the programmers to not break any encapsulation principle by mistake [16]. C++ and C# have very strict rules according the accessibility of class attributes and methods so in both languages the compiler add private access specifier implicitly [16][17]. Apart from private C++ also support public and protected access specifier [19] but in order to use these access specifier programmer must tell the compiler explicitly. C# and Java also support all the access specifier provide by the C++ except the C# provide some more levels of security by adding internal, protected internal, private protected we will discuss these modifiers later in naming encapsulation. Java and Python have slightly less strict rules about default accessibility of attributes and method in Java all the attributes and methods are package scoped [15] as shown in Figure 3 (Code Listing 2) but Java,C#,C++ puts only restriction on the name of attribute but not on reference this problem refers as representation exposure problem Figure 4 (Code Listing 3) [16] and in Python there is no access specifier so data and methods are considered as public. Python hide the details of data and methods by using single and double underscore before the name of attribute and method. Single underscore tells us that it is for internal use only and double underscore changed the name of variable internally and prefixed it with class name which is called name mangling. The variable of method still works outside the class but it requires extra work [20] Figure 3 (line 5,6 Code Listing 1) shows the name mangling in Python.

| Code Listing 1: Name Mangling in Python | Code Listing 2: Package Private in Java |
|---|---|
| 1. **class** Hello:<br>2.   **def** \_\_init\_\_(**self**):<br>3.     **self.**\_\_number =<br>4.     12345<br>5. obj = Hello()<br>6. print(obj.\_\_number)<br>7. #'Hello' object has no<br>8. attribute '\_\_number'<br>9. print(obj.\_Hello\_\_num<br>10. ber) #prints 12345 | 1. **class** Hello{<br>2.   **int** number = 12345;<br>3. }<br>4. **class** HelloWorld<br>5. {<br>6.   **public static void**<br>7.   Main(){<br>8.   Hello obj = new Hello();<br>9.   **System.out.print**(obj.nu<br>10. mber);//`prints 12345}<br>11. } |

*Figure 3. Name Mangling in Python and Package Private in Java.*

**Code Listing 3**: Representation Exposure in Java

1.   **class** Engine {
2.     **private int** engine_no = 10;
3.     **public void** setEngine (**int** e) {
4.     **this**.engine_no = e;
5.     }
6.     **public void** printengine() {
7.     **System.out.println**(**this**.engine_no);
8.     }
9.   }
10. **class** Car{
11.   **private** Engine objEngine;
12.   **public** Car() {
13.     **this**.objEngine = **new** Engine();
14.     **this**.objEngine.setEngine(1000);
15.   }**public** Engine GetEngine() {
16.     **return this**.objEngine;
17.   }**public void** print() {
18.     **this**.objEngine.printengine();}
19. }**public class** Main {
20.   **public static void** main(**String**[] args) {
21.     Car objCar = **new** Car();
22.     Engine objEngine = objCar.GetEngine();
23. }}

*Figure 4.Reference invariant in Java*

**Properties:** To set and get the value of not accessible attributes in Java and C++ languages programmer write the public accessor and mutator functions in C++ and Java the issue with these functions is our classes. But in some situations generally this is not the good practice to write accessor and mutator for every attribute because it can lead to un predictable changes to class attributes Figure 5 (Code Listing 4) shows the simple accessor mutator in C++ [18][19]. C# provide the more dynamic and more encapsulated way of accessing and storing the values of hidden class attributes. C# provide public properties for manipulating hidden attribute values although programmer can still access the attributes by using traditional accessor mutator approach but C# helps to reduce boilerplate code. C# properties is just a simplified version of get and set methods but they enforce the state of encapsulation programmer can write custom business logic as well in predefined properties.

These properties can also be declared as read only and write only properties Figure 6 (Code Listing 5) (line 4,8) shows the working of properties with some custom checks in C# [17]. Python also support properties and provide some level of encapsulation but one can still access the attributes without using properties even class have property for that attribute Figure 7 (line 4,15 - Code Listing 6).

**Auto implemented properties**: If a programmer doesn't want to write any custom logic C# provide mechanism of auto implement properties to avoid extra code and maintain the state of encapsulation as well the auto implement properties secretly define a private backing fields Figure 6 (line 4,6 - Code Listing 5) shows properties with business rules and auto implement properties are shown in Figure 8 Figure 8. Auto Implemented Properties in C#(Line 2 Code Listing 8) [17]. Python also provides the more concise way of declaring properties by using decorators Figure 9 (line 4,9 - Code Listing 8) shows the get and set properties using decorators [20].

**Code Listing 4**: Accessor and Mutator in C++

1.   **class** Hello{
2.     **public** :
3.       **int** getNumber(){**return this**->number;}
4.       **void** setNumber(**int** n){**this**->number = n;}
5.     **private**:
6.       **int** number;
7.   };
8. **int** main()
9. {  Hello obj;
10.   obj.setNumber(25); // Mutator
11.   **cout**<<obj.getNumber(); // Accessor
12.   **return** 0;}

*Figure 5. Accessor and Mutators in C++.*

**Code Listing 5**: Encapsulation Using Properties in C#

1.   **class** Hello{
2.     **private int** number;
3.     **public int** Number{
4.       get { **return** number; }
5.       set { if(value > 0)
6.         {number = value;}
7.       }
8.     }
9. }
10. **class** HelloWorld {
11.   **static void** Main() {
12.     Hello obj = new Hello();
13.     obj.Number = 25; // set data by using set property
14.     obj.Number = -15; // sets nothing
15.   }
16. }

*Figure 6. Properties in C#.*

**Code Listing 6**: Encapsulation using properties in Python

```
1.  class Hello:
2.      def __init__(self):
3.          self.__number = None # semi private attribute
4.      def _set_number(self, number):
5.          self.__number = number
6.      def _get_number(self):
7.          return self.__number
8.      Number = property(_get_number, _set_number)
9.  # main
10. obj = Hello()
11. obj.Number = 25 # setting value using name property
12. print(obj.Number) # getting value using property
13. obj._Hello__number = 10
14. print(obj._Hello__number) # still we can access attribute
15. print(obj.Number) # prints 10
```

*Figure 7. Properties in Python.*

**Code Listing 7**: Encapsulation Using Auto Implemented Properties in C#

```
1. class Hello{
2.     public int Number {get;set;} // auto implement properties
3. }
4. class HelloWorld {
5.     static void Main() {
6.         Hello obj = new Hello();
7.         obj.Number = 25;
8.     }
9. }
```

*Figure 8. Auto Implemented Properties in C#*

**Code Listing 8**: Encapsulation using Decorators properties in Python

```
1.  class Hello:
2.      def __init__(self):
3.          self.__number = None # semi private attribute
4.      @Number.setter
5.      def Number(self, number):
6.          self.__number = number
7.      @property
8.      def Number(self):
9.          return self.__number
10. # main
11. obj = Hello()
12. obj.Number = 25 # setting value using name property
13. print(obj.Number) # getting value using property
14. obj._Hello__number = 10
15. print(obj._Hello__number) # still we can access attribute
    print(obj.Number) # prints 10
```

*Figure 9. Properties using Decorators in Python.*

**Nested classes:** A class have another class in its scope which is called inner class or nested class. The outer class treats the inner class just like other attributes types the access level of inner class also same as other attributes we can declare it with available access modifiers. Inner class can access its outer class attributes even the private attributes [17]. All these selected languages support the nesting of classes. Java support two types of inner classes all of them is hidden from packages and other classes every inner class has reference to its outer class and the object of inner cannot be created without the reference of outer class the Figure 10 (Code Listing 10) [15].

Java also supported anonymous classes these classes always inside the body of another class anonymous class implement interface or extends class and override their behavior anonymous classes in Java shown in Figure 11 (Code Listing 11) [21]. Python also supports inner classes but slightly different from other languages in Python the properties which is associated with object denote by special keyword self and inner class won't have access to its outer class variables as shown in Figure 10 (Code Listing 9) [20].

C++ and C# make details private by default of nested class but C++ and C# both support the concepts of pointers but in order to use pointer in C# we need tell the compiler explicitly by enabling the unsafe mode. In case of reference types in C# the memory used is heap which managed by garbage collector. It is possible that we are accessing the memory which is no longer exist to solve this problem C# provide fixed keyword don't change the address of variable during execution [17]. C# and C++ both support the concept of pointers and we can directly manipulate the memory by using pointers.

The use of pointers in inner can lead to serious security concern the data which is hidden from the outside world is easily changed and access using inner class Figure 12 (Code Listing 12) and Figure 13 (Code Listing 13) shows the working of pointer in inner class and its harmful effects the hidden attribute is access and changed using pointer nested classes of Python and Java are uniformed then nested classes of C++ and C#. Now we will evaluate encapsulation in abstract data types by using our defined qualitative values. The evaluation of each parameter in encapsulation of abstract data types is shown in the Table 2.

| **Code Listing 9**: Nested Classes in Python | **Code Listing 10:** Nested Classes in Java |
|---|---|
| 1. class Outer: | 1. class Outer{ |
| 2.   age = 10 | 2.   private String outer= |
| 3.   def __init__(self): | 3.   "Outer"; |
| 4.     self.name = "outer" | 4.   public Outer(){ |
| 5.   class inner: | 5.   System.out.print("Out |
| 6.     def __init__(self): | 6.   er Created"); |
| 7.       self.inner_name | 7.   } |
| 8.       ="inner" | 8.   class Inner{ |
| 9.       print(name) | 9.   private String inner = |
| 10. #error | 10.   "Inner"; |
| 11. name is no defined | 11.   public Inner(){ |
| 12. outer = Outer() | 12.   System.out.print(outer |
| 13. print(outer.name) # | 13.   ); |
| 14. prints | 14.   // Accessible }} |
| 15. outer | 15. } |
| 16. obj = Outer.inner() | 16. public class Main |
|  | 17. {public static void |
|  | 18. main(String[] args) { |
|  | 19. Outer.Inner obj = |
|  | 20. newOuter().new |
|  | 21. Inner();} |
|  | 22. } |

*Figure 10. Nested Classes in Java and Python.*

**Code Listing 11**: Nested Classes Using Anonymous Class in Java

1. **class** PrintMessage{
2. **public void** printmsg() {
3. **System**.**out**.**println**("Hello Java");
4. }
5. }
6. **public class** Main {
7. **public static void** main(**String**[] args) {
8. PrintMessage objMessage = **new** PrintMessage() {
9. **public void** printmsg() {
10. **System**.**out**.**print**("Overriding by using anonymous class");
11. }};objMessage.printmsg();}}

*Figure 11. Anonymous Classes in Java.*

**Code Listing 12**: Harmful Nested Classes in C++

| | |
|---|---|
| 1. **class** Outer{ | 14. **int** main() |
| 2. **private** : | 15. { |
| 3. **int** OuterNumber = 10; | 16. Outer :: Inner obj; |
| 4. **public**: | 17. Outer oob; |
| 5. **class** Inner{ | 18. **int** *ptr |
| 6. **int** *ptr; | 19. =obj.getptr(&oob); |
| 7. **public**: | 20. *ptr = 25; |
| 8. **int** *getptr(Outer *obj){ | 21. **return** 0; |
| 9. **this**->ptr = | 22. } |
| 10. &(obj>OuterNumber); | |
| 11. **return this**->ptr; | |
| 12. } | |
| 13. };}; | |

*Figure 12.Security Issue in Nested Classes of C++.*

**Code Listing 13:** Harmful Nested Classes in C#

| | |
|---|---|
| 1. **class** Outer{ | 15. } |
| 2. **private int** x=90; | 16. **class** HelloWorld |
| 3. **public unsafe class** | 17. { |
| 4. inner{ | 18. **static void** Main() |
| 5. **int*** ptr; | 19. { |
| 6. **public int** * | 20. **unsafe** |
| 7. method(Outer obj) | 21. { |
| 8. { | 22. Outer outer = **new** |
| 9. **fixed** (**int** *p = &obj.x) | 23. Outer(); |
| 10. { | 24. Outer.inner obj2 = **new** |
| 11. ptr = p; | 25. Outer.inner(); |
| 12. } | 26. **int** *p = |
| 13. **return** ptr; | 27. obj2.method(outer); |
| 14. } | 28. *p = 45;// value of x is |
| | 29. now 45 |
| | 30. }}} |

*Figure 13. Nested classes in C#.*

*Table 2. Encapsulation in Abstract Datatypes.*

| Languages | Implicit Access Modifier | Properties | Auto Implemented Properties | Nested Classes |
|---|---|---|---|---|
| **C++** | Fully | No | No | Partially |
| **Java** | Partially | No | No | Fully |
| **Python** | No | Partially | Mostly | Mostly |
| **C#** | Fully | Fully | Fully | Partially |

b) NAMING ENCAPSULATION

Large programs written in the form of libraries and large modules which is managed by different developers. It is important that module provide by language also have an some kind of security so that each module and library have separate identity in terms of names each have its own unique name this concept is known as naming encapsulation [15]. In this research we evaluate the concept of naming encapsulation by following parameters (I) package access only (II) friend classes (III) definition in multiple files.

**Package access:** Java and Python provides the mechanism to group the logically related classes by providing packages in Python a single python file is called module and collection of these modules is package. Java support package only by making the class the final class in Java cannot instantiate outside the package [15] in Figure 14 (Code Listing 14) shows the working of final keyword before class inside a package. Packages in Python is created by empty special file named init__ with .py extension when the package is imported the init.py is executed by default we access every module in package if we want to restrict something then we need to define __all__ variable and define those module and variable in the list so that if someone import all then only allowed module this also provides the security of variable and modules only[22] shows even with * import only those modules which are allowed in __all__ list in init.py will be import but in Figure 15 (Code Listing 15) if we remove line 2 then line 6 will generate error although python provides some kind of package restriction but not like Java.

C++ and C# provide this facility by using namespaces. The logically related classes are the part of same scope and members of same namespace can refer each other without any special notation[19]. When using global imports from namespace it might be possible that both namespace contain same name classes over use of global imports can lead to name clashes name the namespace in C++ have access modifiers so everything is accessible [19] Figure 16 (Code Listing 16) showing the namespace working namespace in C++. C# have almost similar namespace concept like C++ but only difference in the accessibility of members of namespace. C# provide internal and public by default every member of namespace is public [17].

**Friend classes:** Friend classes and functions is a very controversial topic in programming because it violates the encapsulation and change the internal state of object by accessing private field but without support of friends many real-world application interfaces are nearly impossible to implement [23][24]. C++ is the only language which support the concept of friends explicitly friend classes and function in C++ is shown in Figure 17 . Java implements the friend's mechanism by using package private variables. In Java classes of same package are partial friends of one another but the classes of another package cannot directly access the package private attributes this is the Java way of implementing the friends behavior because logically related classes will be part of same package [23]. In C# we achieve friendship by using keyword internal before attribute function or class name the internal keyword in C# is very similar to Java package access the internal keyword works inside assembly classes of same assembly can access their internal attributes and methods Figure 18 have two different namespaces with two different classes but due same assembly they share internal attribute [24]. As we already discussed in abstract datatype encapsulation everything is public in Python so there is no need of such mechanism.

**Definition in multiple files:** While working with large classes we often change or update the code written in functions to solve this problem C++ and C# languages provide the mechanism to partition classes into multiple files [17], [18]. C++ allows programmer to write separate specification and implementation by using header files and cpp files the Figure 19 (Code Listing 19) shows the class have separate files one is class file and other is implementation file [18]. C# uses the concept of partial keyword to partition the classes and functions. Partial class or functions only observe during development process but at compile time the compiler merge all the chunks from different files and composed into a single class or function Figure 20 shows the working of partial classes and methods using two different files [17]. Java does not support stand-alone functions and writing methods outside the class parenthesis because of object oriented behavior [15]. Python do not support partial function and separate implementation of class of method outside the class the naming evaluation is shown in Table 3.

**Code Listing 14**: Package Access Classes in Java

| Main Class | Instrument Class |
|---|---|
| 1. **import** Music.Guitar;<br>2. **import** Music.*;<br>3. **public class** Main {<br>4. **public static** void<br>5. main(**String**[] args) {<br>6. Guiter objGuitar = **new**<br>7. Guitar();<br>8. InstrumentString<br>9. objString = **new**<br>10. InstrumentString();<br>11. //error The type<br>12. InstrumentString is not<br>13. visible<br>14. }<br>15. } | 1. **package** Music;<br>2. **final class**<br>3. InstrumentString<br>4. {<br>5. **int** string_count;<br>6. **String** string_type;<br>7. **Public**<br>8. InstrumentString()<br>9. {<br>10. **System.out.print**("Stri<br>11. ng Class");<br>12. }<br>13. } |

*Figure 14. Package Restriction on Classes in Java*

**Code Listing 15**: Packages in Python

1. **from** Music **import** *
2. **from** Music **import** InstrumentString
3. guiter = Guiter.Guiter()
4. banjo = Banjo.Banjo(3,"steel")
5. int=InstrumentString.InstrumentString(3,"Steel") #
6. NameError:
7. name 'InstrumentString' is not defined
8. #code inside __init__.py
   __all__ = ["Guiter", "Banjo"]

*Figure 15. Package access only example in Python*

**Code Listing 16**: Namespaces in C++

| | |
|---|---|
| 1. **#include** <iostream><br>2. **#include** "Music.h"<br>3. **using namespace std**;<br>4. **using namespace**<br>5. Music;<br>6. **int** main()<br>7. {<br>8. Guitar guitar;<br>9. Banjo banjo;<br>10. InstrumentString<br>11. intrument_string;<br>12. **return** 0;<br>13. } | 14. **#ifndef**<br>15. **MUSIC_H_INCLUD**<br>16. **ED**<br>17. **#define**<br>18. **MUSIC_H_INCLUD**<br>19. **ED**<br>20. **namespace** Music {<br>21. **class** Guitar{/../<br>22. };**class** Banjo{/../<br>23. };<br>24. **class**<br>25. InstrumentString{/../<br>26. };<br>27. }<br>28. **#endif** |

*Figure 16. Global Namespace in C++.*

**Code Listing 17**: Friend Classes in C++

1. **class** A
2. {
3. **int** x = 90;
4. **friend class** B;
5. };
6. **class** B
7. {
8. **public**: **void** method(){
9. A a ;
10. a.x = 15; // valid
11. }
12. };

*Figure 17. Working of Friends Functions and Classes in C++.*

**Code Listing 18**: Friend Classes in C#

| | |
|---|---|
| 1. **using** System;<br>2. **using** A;<br>3. **class** HelloWorld {<br>4. **static void** Main() {<br>5. A obj = **new** A();<br>6. obj.msg = "Hello C#";<br>7. }<br>8. } | 1. **namespace** A{<br>2. **class** A{<br>3. **Internal String** msg<br>4. ="Hello world";<br>5. }} |

*Figure 18. Implementation of Friends using C#.*

**Code Listing 19**: Separate implementation in C++

| Specification file | Implementation file |
|---|---|
| 1. **class** Test { | 1. **#include** |
| 2. **public**: | 2. "HelloWorld.h" |
| 3. **void** printString(); | 3. **void** |
| 4. **void** printNumber(); | 4. Test**::**printString(){ |
| 5. }; | 5. **cout**<<"Print String"; |
| | 6. } |
| | 7. **void** |
| | 8. Test::printNumber(){ |
| | 9. **cout**<<"12"; |
| | 10. } |

*Figure 19. Partition of Class in C++.*

**Code Listing 20**: Partial Classes in C#

| | |
|---|---|
| 1. // partial class .cs2 | 1. **public partial class** |
| 2. **public partial class** | 2. PartialClass { |
| 3. PartialClass { | 3. **private int** num; |
| 4. **partial void** | 4. |
| 5. partialMethod (){ | 5. **public** |
| 6. | 6. PartialClass( |
| 7. **Console.Writeline** | 7. **int** t) |
| 8. (' | 8. { |
| 9. partial method');} | 9. **this**.num = t; |
| 10. } | 10. } |
| | 11. **public partial void** |
| | 12. partialMethod();} |

*Figure 20. Partial Classes and Methods in C#.*

*Table 3. Evaluation of Naming Encapsulation.*

| Language | Package Access | Friend Classes | Definition in Multiple Files |
|---|---|---|---|
| C++ | No | Fully | Fully |
| Java | Fully | Mostly | No |
| Python | Mostly | No | No |
| C# | Partially | Mostly | Partially |

### 2) RELATIONSHIPS AMONG OBJECTS

In our daily life the one object is composed of many other objects to use the services of each other. In object-oriented programming an object also contains other objects or used services of another objects the relationship among object also knows as associations. An object which is composed of another object it is refers to as aggregation it is also refers as 'has a' or 'whole part' relationship [25]. Aggregation have strong form which is called composition the life time of object in composition is dependent on whole life time and it belongs to one whole at a time[3] [20]. In this section we are going to evaluate aggregation and composition by using the following parameters I) uniform access II) implicit initialization of attributes III) initialization of constant attributes IV) constructor overloading V) copy constructor VI) automatic memory management. We will look how each language define aggregation and composition.

**Uniform access:** The uniform principle states that "All services offered by the modal should be available through notation" the uniformity in access provides consistency [26]. The uniform access can strongly effect on language because if someone is not familiar with all the constructs provided by the languages chances are misuse of features and it will take more time to solve even simple problems [15].

C++ allow both value types and reference type objects the class member access operator is different for each type [3] [15]. Value type objects access members by using dot operator and reference type variable used to access members by using hyphen and greater than sign although we access members by using dot notation in dereference approach but is not elegant and due precedence issues [3] in Figure 21 (Code Listing 21) class C have composite and aggregated objects and both call their respective method differently. By default, C# shows uniformity in case of value types and reference types but because of C# support pointers then like C++ C# used arrow notation to access methods and attributes.

Java and Python provide more uniformity than C++ and C#. Both languages have only one way to instantiate class and only one way to access member of the class. Java uses new keyword to create object on heap in Figure 22 shows the composition and aggregation in Java the syntax is similar but initialization of object is different the composite object is initialize inside the class and in aggregation only the reference of object is passed [20] [21]. Python uses the simple dot notation to access methods and class attributes in composition and aggregation in python have very clear and uniform code just like Java except new keyword [22].

**Code Listing 21**: Uniform Access in C++

```
1.  class C{
2.  private :
3.      A Aobj;
4.      B *Bobj;
5.  public: void Cmethod (){
6.      Aobj.Amethod();
7.      Bobj = new B();
8.      Bobj->Bmethod();
9.    }
10. };
```

*Figure 21. Dot and Pointer to Object Member Access using Composition in C++.*

**Code Listing 22**: Uniform Access in Java

```
1.  class C{
2.  private
3.     A Aobj = new A();
4.     B Bobj;
5.  public void method (B obj){
6.     Aobj.Amethod();
7.     Bobj = obj;
8.     Bobj.Bmethod();}};
```

*Figure 22. Access of Members in Composition and Aggregation in Java.*

**Implicit initialization :** Default initialization is another very important feature provide by the language because if language does not have proper way to initialize the attributes of class then program is more prone to run time exception or ambiguous behavior so it's always a good practice that every attribute of class should be initialized by a valid value and to hold a class invariant a condition that must be hold before calling the method [18] [19] .

C++ do not support any default mechanism to initialize the class attributes so the class attributes always contain the garbage value in Figure 23 (Code Listing 23 line 3,4) to pointer objects in class one with garbage value and other is properly initialize to valid value it is useful when maintaining checks. To initialize attributes in C++ we have two methods one is assign values in constructor and the other in place member initialization before C++11 in Figure 23 due to default value of nullptr we can easily make our code more secure and reliable if class contain pointer then it's necessary that we need to check that pointer points to valid address [18].

Python don't have any implicit mechanism to initialize attributes but it forces programmer to initialize with some value to variables and objects otherwise it is a compile time error. In order to assign well defined values to attributes we have to tell compiler explicitly by using self-keyword either in initializer which call automatically on object creation or a class method which we call explicitly and assign with some initial value shows the initialization of attributes using class method or initializer [20] [22]. Java provide implicit initialization of attributes according to datatypes of attributes if we call method on object which is null reference if we don't initialize the valid values for objects then we will get the run time exception of null pointer so we need to be careful while implementing composition and aggregation Figure 24 shows the objects initialize with null default value and integer variable with zero [21].

C# initialized class fields automatically just like Java with appropriate values. We can write constructor for initializing the fields as well. C# provide initialization of values using default constructor like Java C++. C# auto properties are not nullable by default so if programmer wishes to assign null as a default value, then puts a question mark before property name otherwise default value is Compulsory to initialize with some value other than null. if class has multiple constructors, then it's difficult to know all of the constructors. C# provide the object initializer syntax this syntax initializes the public properties and public fields before calling the constructor Figure 25 (Code Listing 25) shows the object initializing syntax[17].

**Code Listing 23**: Implicit initialization in C++

```
1.  class C{
2.  private :
3.     A *Aobj;
4.     B *Bobj = nullptr;
5.  public:
6.     void method (){
7.       if(!Bobj){
8.         Bobj = new B();
9.       }else{
10.        Bobj->Bmethod();
11.      }
12. }};
```

*Figure 23. Garbage value issue and In-place initialization in C++.*

**Code Listing 24**: Implicit initialization in Java

```
1.  class C{
2.    private
3.      int num;
4.      A Aobj;
5.      B Bobj;
6.  };
```

*Figure 24. Default values of attributes in Java.*

**Code Listing: 25** Object Initializer in C#

```
1.  class C
2.  {
3.    public A Aobj { set; get; }
4.    public B Bobj { set; get; }
5.  }
6.  class test
7.  {
8.    static void Main()
9.    {
10.     B bobj = new B();
11.     C obj = new C(){ Aobj = new A() , Bobj = bobj};
12.   }
13. }
```

*Figure 25. Object initializer in C#.*

**Initialization of constant attributes:** Constants attributes are those attributes which don't change their value throughout the program execution. If certain fields of class don't change during program execution, then it is good practice to set them as constant in order to follow the principle of least privileged constant fields are less error prone [27].A simple way is to initialize constant filed in class is with any value while declaration which is commonly known as in place member initialization [18]. But some cases. We may need to initialize the constant fields during runtime or with users' value. In order to solve this problem some programming language provide such constructs.

C++ provide the concept of member initializing lists the values of member initializing lists are set before constructor call Figure 26 (Code Listing 26) presents the working of constant fields and working of member initializing list in C++ [18]. Java provide in place member initialization of final field if it is not initialized while declaration then it must be initialized in every constructor of class [27]. it's the Java way of declaring the constant fields because Java don't support any mechanism Figure 27 (Code Listing 27) shows the initialization of final fields in Java. Python don't support any type constant so it's not possible to create constant objects in Python.

C# support traditional C++ keyword constant but unlike Java C# don't allow constant to initialize in the constructor or don't have any mechanism like member initializing list infect we cannot use const keyword before creating instance of any class. C# also support another keyword readonly the difference between constant and readonly is if value is known at compile time, then we used constant and if the value is unknown then we readonly Figure 28 (Code Listing 28) [17]. C# also support two more keywords first is 'required' and second is 'init' but required keyword is only support version greater than C# 10. 'init' is useful when we are working with immutable fields and required enforce that property or field must be initialize when object is created [28].

**Code Listing 26**: Initializing List in C++
1. **class** A {
2. **const int** field1 = 15;
3. **public** : **const string** field2;
4. **public** :
5. A(**string** x) : field2(x){
6. }};
7. **class** B{
8. **const** A *ptr;
9. public : B( **const** A &aobj): ptr(&aobj){
10. }};
11. **int** main()
12. {
13. **const**  A aobj("constant string");
14. B bobj(aobj);
15. }

*Figure 26. Initialization of Constant Members in C++.*

**Code Listing 27**: Initialization of Final Instance in Java
1. **class** A {
2. **final int** number = 15;// won't change
3. **public** A(){
4. //**this**.number = 8; // run time error
5. }
6. }
7. **class** B {
8. **final String** name;
9. **final** A Aobj;
10. **public** B(){
11. **this**.name = "setting name";
12. **this**.Aobj = **new** A();}
13. }

*Figure 27. Initializing of Final Members in Java.*

**Code Listing 28**: Constant and Readonly Fields in C#
1. **class** A{
2. **readonly int** number;
3. **const int** NUMBER = 123;
4. **public** A(**int** n){
5. number  = n;// can be changed in constructor
6. }
7. }
8. **class** B{
9. **readonly** A obj;
10. public B(){
11. obj = **new** A(25);}
12. }

*Figure 28.Working const and readonly fields in C#.*

**Constructor Overloading:** We have already discussed constructors and their roles. Just like function overloading some languages support constructor overloading the main idea behind the concept is to provide flexibility to create an object with multiple options

Java C# and C++ support constructor overloading the concept of constructor overloading is similar in these three languages Figure 29 shows constructor overloading in Java [18]. Python have a constructor and initializer but constructor in Python is rarely used the Python constructor is called new which creates an object and initializer initialize the object so if want some action before initialization then need to write the constructor Figure 30 shows the constructor in Python [20].

**Code Listing 29**: Overloading of Constructor in Java

```
1.  class B
2.  {
3.  private A Aobj;
4.  public  B(){
5.  this.Aobj = A();
6.  }
7.  public B(A obj) {
8.  this.Aobj = obj;
9.  }
10. }
11. public class Main {
12. public static void main(String[] args) {
13. B obj1 = new B();
14. B obj2 = new B(new A());       }}
```
*Figure 29.Overloading of Constructor in Java.*

**Code Listing 30**: Constructor in Python
```
1.  class MyClass:
2.  def __new__(cls, *args, **kwargs):
3.  obj = super().__new__(cls)
4.  obj.name = "hello Python"
5.  print("iam constructing object")
6.  return obj
7.  object = MyClass()
```
*Figure 30.Constructor in Python.*

**Copy constructors:** Every object-oriented language provides some way of cloning one object to another object without changing it. The cloning is done by special constructor which is called copy constructor. While referencing other object into our classes there is two common security issue can occur first is performing shallow copies instead of deep copies and second is return original objects instead of their copies default copy constructor in C++ and C# performs member by member copy. [18], [21].

To solve the above problem, we forced to override the default behavior of copy constructor. Copy constructors can change data because the attributes are accessible in copy constructors to avoid modification of data it is usually a good practice to specify constant before object reference in argument in Figure 31 we are passing the copy of A class object not the original object in C++. Java also provide the copy constructor same as C# in Figure 32 we are returning copy of original object not the original object [18]. We can also disable the coping outside the class by declaring copy constructor private [3].

Python don't have any copy constructor in order to copy objects we need to write user defined method to copy contents and call it as a simple class method Figure 33 (Code Listing 33 line 7,9) illustrate the copy of object in Python. Beside copy constructor some programming languages provide some cloning functions these cloning functions help to replicate the given object except older languages like C++ mostly new languages support cloning methods Java C# and Python support these cloning methods the cloning methods don't call copy constructor the cloning functions do shallow copy by default so we need aware with that. In order to use clone method we need implement the clonable interface in Java and Icloneable in C# and in Python we have to import copy module[22][27][28].

**Code Listing  31**: Custom Copy Constructor in C++
```
1.  class A {
2.  string field2="String";
3.  public :
4.  A(){}
5.  A(const A &obj){
6.  this->field2 = obj.field2;}
7.  };
8.  class B{
9.  const A *ptr;
10. public  : B( const A  &aobj): ptr(&aobj){}
11. };
12. int main()
13. {
14. A aobj;
15. A aobj2 = aobj;
16. B bobj(aobj2);
17. }
```
*Figure 31. Custom Copy Constructor in C++.*

**Code Listing 32**: Custom Copy Constructor in Java
```
1.  class A {
2.  int number = 15;
3.  public A(){}
4.  public A(A obj){
5.  this.number = obj.number;}
6.  }
7.  class B {
8.  private A Aobj;
9.  public B(){
10. this.Aobj = new A();
11. }
12. public  A getAObject(){
13. return new A(this.Aobj); }
14. }
15. public class Main
16. {
17. public static void main(String[] args) {
18. B b = new B();
19. A aobj = b.getAObject();}}
```
*Figure 32. Custom Copy Constructor in Java.*

**Code Listing 33**: Custom Copy Method in Python

1. **class** A:
2. **def** \_\_init\_\_(**self**):
3. **self**.num = 12;
4. **def** copy(**self**, aobj):
5. **self**.num = aobj.num
6. **class** B():
7. **def** \_\_init\_\_(**self**,aobj):
8. **self**.Aobj = A()
9. **self**.Aobj.copy(aobj)

*Figure 33.Custom Copy Function in Python*

**Automatic Memory Management:** Memory management is a very important feature in programming language. Some languages support both stack heap base objects while some languages only support heap-based objects. C++ language support stack based as well heap-based objects the stack-based objects remove from memory when the function call return but heap-based object needs to handle by programmer if a programmer forgets to de allocate the memory the memory leakage issue occur or dangling reference problem can also occur [18]. In heap based objects C++ provide more flexibility than other languages and program execution efficiency but this is prone to critical errors and lead to system failure [23] in Figure 34 (Code Listing 34 line 10) allocate the object dynamically but did not delete. When we allocate memory dynamically in classes its always compulsory that we have to write destructor to delete the memory when object is no longer needed [3]. C++ 11 version introduces the smart pointers. Smart pointers are automatically handled by C++ this helps to prevent the memory leaks to use smart pointers we must include the memory directive Figure 34 (Code Listing 34 line 11) [18].

Memory management in Java is automatically handle by JVM. JVM have a process known as garbage collector which remove those objects which are not referenced by any other object so writing destructor in Java is not a common practice The destructor in Java called finalize method and calling of finalize method can cause problems because finalize method calls the garbage collector to sweep out the unreferenced objects but it is rarely used [23][27]. Python also done automatic garbage collection the objects but by the usage of 'gc' module we can tune the some options of garbage collector e:g enable and disable or access objects found by the garbage collector. Python garbage collector invoked automatically when number of allocations in greater than number of deallocations. Python have destructor which is called delete but it is very dangerous to use it because Python automatic garbage collector ignore the objects which defined the delete method in class Figure 35 (Code Listing 35 line 10) if don't explicitly remove the B class object then automatic garbage collector don't collect it and it will cause a memory leak issue [22]. C# provide CLR garbage collector it works when object is unreachable in codebase. C# also support defining destructors in class while using unsafe code then its programmer's responsibility to managed memory. By using GC class we can call garbage collection mechanism Figure 36 (Code Listing 36) shows the destructor in C# [17]. Evaluation of relationships among objects is shown in Table 4.

**Code Listing 34**: Pointers without Destructor in C++

1. **class** A {
2. **public** :**string** field2="String";
3. **public** :
4. A(){ }
5. };
6. **class** B{
7. A *ptr;
8. **unique_ptr**<A> Aptr;
9. **public** : B( ){
10. ptr = **new** A();
11. **unique_ptr**<A>Aptr (new A());
12. }
13. };

*Figure 34. Memory Leakage and smart pointers in C++.*

**Code Listing 35**: Destructor in Python

1. **class** A:
2. **def** \_\_init\_\_(**self**):
3. **self**.num =20;
4. **class** B:
5. **def** \_\_init\_\_(**self**):
6. **self**.aobj = A()
7. **def** \_\_del\_\_(**self**):
8. **del self**.engine
9. b = B()
10. **del** b

*Figure 35. Implementation of Delete Method in Python.*

**Code Listing 36**: Destructor in C#

1. **class** A
2. { **public  int** field { **init**; **get**; }
3. };
4. **class** B
5. {
6. **private** A aobj;
7. **public** A()
8. {
9.     aobj = **new** A {field = 7500};
10. }
11. ~B()
12. {
13. aobj = **null**;
14. }};

*Figure 36. implementation of Destructor in C#.*

*Table 4. Relationships among objects.*

| Language | Uniform Access | Implicit initialization of attributes | Initialization of constant attributes | Constructor overloading | Copy Constructor | Automatic memory management |
|---|---|---|---|---|---|---|
| C++ | Mostly | No | Fully | Full | Partially | Mostly |
| Java | Fully | Fully | Mostly | Fully | Partially | Fully |
| Python | Fully | No | No | No | No | Fully |
| C# | Fully | Fully | Fully | Fully | Fully | Fully |

*3) Generalization*

We can define specialized objects in terms of more general objects [18]. Inheritance allows us to create generalized classes and by using these generalized classes we create specialized classes in our programs [18]. In previous section we discussed the 'has a' relationship among classes now in this section we will be discussed the 'is a' relationship in classes in object-oriented programming the inheritance is used to create the 'is a' relationship among classes. The child class have all characteristic of its parent class as well as some of its own [18].The usage of inheritance reduced the size of the program by using code reusability.

In this study we will evaluate the concept of inheritance by using the following parameters (I) support base class access specification (II) support multiple inheritance (III) support non-extendable classes.

**Support base class access specification:** Base class access specification controls the accessibility of the inherited members in derived class. C++ supports three types of access specification public, private, protected and if no access specification is provided then it is private by default. The base class speciation provide flexibility and become the filter of members which is not needed in Figure 37(Code Listing 37) illustrate the public functions of base class is become private when inheritance is private and protected in case of protected inheritance [18].

Java support only public access specifier before class name we can only restrict extra restriction on attributes of class no on whole class. Java use extend keyword before class name to inherit any class the accessibility of inherited members depends on the access specifier in base class. Java protected members can be access by subclass even if the subclass belongs to different package but package access can only be visible inside the package Figure 38 (Code Listing 38) shows the Java default access level inheritance in which we can access protected methods in same package [21]. In Python every class is subclass of super class object so we can implement some of the object class methods. Python do not support access specifier in inheritance all we have to do is simple pass a class name in parenthesis [20]. As we already discussed that

C# provide more accessibility controls as compared to other programming languages for the sake of clear understanding, we presented the accessibility of base class members in derived class in **Error! Reference source not found.** . Like Java and C# does not provide any base class access specification but it provides the wide range access specifier to class members [17].

**Multiple inheritance:** Multiple inheritance is the most controversial type of inheritance. It is always good practice to not use multiple inheritance unless it is really required. Multiple is just a class having two more than one base classes at the same time this leads to several ambiguities [3][20].

C++ and Python both support multiple inheritance. Multiple inheritance very much prone have ambiguity in code. The problem a programmer is faced is ambiguity in function calling if both the base classes have function with same name compiler would not know which function to call. C++ resolve this ambiguity by using scope resolution with class name before method call Figure 39 (Code Listing 39 line 15,18) illustrate the ambiguity in function call on line 16 but to resolve this error use scope resolution operator by class name [19].

Python provide the different approach to solve this error it maintains a list by more specialized class to least specialized class by default Python calls the most specialized base class method if other base classes have method with same name Figure 40 (Code Listing 40 line 10,17) illustrate the both ways of calling the super class method with same name. Derived class with more than one base class ordering is determined by C3 linearization algorithm. Sometime a situation can occur that algorithm can't make the proper ordering Figure 40 (Code Listing 41) illustrate the inheritance with complex order and Python won't find a consistent resolution and throws an exception that's why it's always a good practice to avoid multiple inheritance [22].

The second problem in multiple inheritance arises when one derived class have more than one base class and those base classes have same parent class Figure 41 (Code Listing 42) illustrate the diamond shaped inheritance in C++ [3]. C++ provide the solution for diamond problem by doing virtual inheritance Figure 42 (Code Listing 43 line 5,10) illustrate the working of virtual inheritance. By adding virtual keyword before base class C++ makes a single copy of every virtual base class for derived by doing so compiler have only one copy of method [19]. Diamond problem won't create much confusion in Python as in C++ because Python calls method in more specialized base class so it won't go for base class Figure 43 (Code Listing 44) illustrate the diamond inheritance but Python calls more specialized base class method [22]. Java and C# support only one direct base class at a time but for implementing complex hierarchies they introduced the concept of interfaces which we will discuss in the next section[17].

**Code Listing 37**: Base Class Access Specification in C++

1. **class** Base {
2. **public** :
3. **void** func1(){}
4. **protected** :
5. **void** func2(){}
6. **private**: **void** func3() {}
7. };
8. **class** Derived : **public** Base {};
9. **int** main()
10. {
11. Derived obj;
12. obj.func1();// accessible
13. obj.func2();//accessible in derived class
14. }

*Figure 37. Base Access Specification in C++.*

**Code Listing 38**: Access Specification in inheritance in Java

1. **class** Base {
2. **public**
3. **void** publicMethod() {
4. }
5. **protected void** protectedMethod() {
6. }**private void** privateMethod() {
7. }}
8. **class** Derived **extends** Base {
9. **void** derivedMethod() {
10. protectedMethod();}
11. }
12. **public class** Main {
13. **public static void** main(**String**[] args) {
14. Derived objDerived = **new** Derived();
15. objDerived.derivedMethod();
16. objDerived.protectedMethod();
17. }}

*Figure 38.Base Class Access Specification in Java.*

**Code Listing 39**: Ambiguous Functions calls in C++

1. **class** Base1 {
2. **public**:
3. **void** printMsg(){
4. **cout**<<"Hello from Base 1";}
5. };
6. **class** Base2 {
7. **public**:
8. **void** printMsg(){
9. **cout**<<"Hello from Base 2";}
10. };
11. **class** Derived : **public** Base1 , **public** Base2 {
12. };
13. **int** main()
14. {
15. Derived obj;
16. obj.printMsg();
17. obj.Base1::printMsg();
18. obj.Base2::printMsg();
19. **return** 0;}

*Figure 39. Ambiguous Function Call in Multiple Inheritance in C++*

.

| **Code Listing 40**: Method Resolution Order in Python | **Code Listing 41:** Ambiguous Method Resolution in Python |
|---|---|
| 1. **class** Base1: | 1. **class** Base1: |
| 2. **def** method(**self**): | 2. **def** method(**self**): |
| 3. **print**("base 1") | 3. **print**("base 1") |
| 4. **class** Base2: | 4. **class** |
| 5. **def** method(**self**): | 5. Base2(Base1): |
| 6. **print**("base 2") | 6. **def** func(**self**): |
| 7. **class** | 7. **print**("base 2") |
| 8. Derived(Base1,B | 8. **class** |
| 9. ase2): | 9. Derived(Base1, |
| 10. **def** | 10. Base2): |
| 11. alternate(**self**): | 11. pass |
| 12. Base1.method(**se** | 12. d = Derived() |
| 13. **lf**) | |
| 14. Base2.method(**se** | |
| 15. **lf**) | |
| 16. d = Derived() | |
| 17. d.method() | |

*Figure 40.Method Resolution Order and Ambiguous Method Resolution in Python.*

**Code Listing 42**: Diamond Shaped Inheritance in C++

**Code Listing 42**: Diamond Problem in C++

```
1.  class Base{
2.  public :
3.  void show(){}};
4.  class A : public Base {
5.  public :
6.  void classAmethod(){}
7.  };
8.  class B : public Base {
9.  public:
10. void classBmethod(){};
11. };
12. class C : public  A , public B{
13. public:
14. void classCmethod()
15. {}
16. };
17. int main()
18. {
19. C obj;
20. obj.show();
21. return 0;}
```

Figure 41.Diamond Problem in C++.

**Code Listing 43**: Virtual Inheritance in C++

```
1.  class Base{
2.  public :
3.  void show(){}
4.  };
5.  class A : virtual public Base {
6.  public :
7.  void classAmethod(){
8.  }
9.  };
10. class B :virtual public Base {
11. public:
12. void classBmethod(){
13. };
14. };
15. class C : public  A , public B{
16. public:
17. void classCmethod(){}
18. };
19. int main()
20. {
21. C obj;
22. obj.show();
23. return 0;}
```

Figure 42. Virtual inheritance in C++.

**Code Listing 44**: Diamond Inheritance in Python

```
1.  class A:
2.  def print(self):
3.  print("A class")
4.  class B1(A):
5.  def printname(self):
6.  print("B1 Class")
7.  class B2(A):
8.  def printname(self):
9.  print("B2 Class")
10. class C(B1,B2):
11. pass
12. c = C()
13. c.printname()
```

Figure 43. Diamond Shaped Inheritance in Python.

**Non-Extendable Classes:** Not every class needs to be extendable as we discussed that only generalized classes are extendable. It is a good practice to make non extendable classes in program when someone is designing the utility classes [17]. Declaring non extendable classes can prevent the subclass errors and prevent from other security reasons [27].

In Java we can create non extendable class by using final keyword. In C++11 final keyword introduces which prevents the class from being inherit same as Java Figure 44 shows the final classes in C++ and Java [18] [27]. C# have sealed keyword to prevent the class from being inherit Figure 45(Code Listing 47 line 3) generate the compile time error [17]. Python do not support non extendable classes. The evaluation of inheritance is shown Table 5.

| **Code Listing 45**: Non-Extendable classes in C++ | **Code Listing 46:** Non-Extendable classes in Java |
|---|---|
| 1. **class**<br>2. NonExtendab<br>3. leClass **final**<br>4. {<br>5. };<br>6. **class** Derived<br>7. : **public**<br>8. NonExtendab<br>9. leClass{<br>10. }; | 1. **final class**<br>2. Base<br>3. {<br>4. }<br>5. **class** Derived<br>6. **extends** Base {<br>7. } |

Figure 44.Final classes in C++ and Java.

**Code Listing 47**: Sealed Classes in C#

```
1. sealed class NonExtendable
2. {
3. }class Derived :NonExtendable
4. {
5. }
```

Figure 45. Non-Extendable Class in C#.

*Table 5.Evaluation of inheritance*

| Language | Base Class Access Specification | Support Multiple Inheritance | Support Non-Extendable Class |
|---|---|---|---|
| C++ | Fully | Fully | Fully |
| Java | Mostly | No | Fully |
| Python | No | Fully | No |
| C# | Mostly | No | Fully |

### *4) Polymorphism and Realization*

Polymorphism means on thing can have several forms [3]. In object-oriented programming polymorphism is achieved by creating the reference of base class which refer the objects of subclass[21]. There are two types of polymorphism (I) Compile time polymorphism and (II) Run time polymorphism [19]. But in this research, we will discuss the runt time polymorphism in detail. We will Evaluate the polymorphism by using the following parameters (I) by default dynamic binding (II) enforce override keyword (III) support prevent the method from overridden (IV) support operator overloading (V) support interfaces.

**Default Dynamic Binding:** Finding the correct binding of method is known as binding [21]. We can create the base class pointer which have reference of child class. If one method with same name exists in base and child class then correct version of method needs to be call. C++ and C# choose the functions during compile time which is called early biding or static binding Figure 46 (Code Listing 48 line 14,16) calls the base class show method although it has reference of derived class and Figure 46 (Code Listing 49 line 20,23). In C++ language we need to specify base class destructor as virtual in order to correctly call the destructor otherwise wrong calling may lead memory leakage problem.

In order to force the compiler to late binding C++ and C# provide the virtual keyword before method name and make the function dynamically bound Figure 47 (Code Listing 50) illustrate the late binding in C# [3] [17] [19] . Java and Python resolve the method at the run time base on the reference type which is called dynamic binding and call the correct version of method using any extra keyword Figure 48 illustrate the example of late binding in Java and Python [27].

| **Code Listing 48**: Static binding in C++ | **Code Listing 49:** static binding in C# |
|---|---|
| 1. **class** Base<br>2. {<br>3. **public** void<br>4. Show()<br>5. {<br>6. }<br>7. }**class** Derived<br>8. :**public** Base<br>9. {<br>10. **public void** show<br>11. () { }<br>12. }<br>13. **int** main(){<br>14. Base *obj = **new**<br>15. Derived();<br>16. obj->show(); | 1. **class** Base<br>2. {<br>3. **public void**<br>4. Show()<br>5. {}<br>6. }**class** Derived<br>7. :**public** Base<br>8. {<br>9. **public void**<br>10. show () { }<br>11. }<br>12. **class** test<br>13. {<br>14. **static void**<br>15. Main()<br>16. {<br>17. Base obj =<br>18. **new**<br>19. Derived();<br>20. obj.Show();<br>21. }} |

*Figure 46.Early Binding in C++ and C#.*

| **Code Listing 50**: Virtual Methods in C# |
|---|
| 1. **class** Base<br>2. {<br>3. **public virtual void** show()<br>4. {<br>5. **Console**.**WriteLine**("base");<br>6. }<br>7. }**class** Derived :Base<br>8. {<br>9. **public override  void** show() {<br>10. **Console**.**WriteLine**("child");<br>11. }<br>12. }**class** test<br>13. {<br>14. **static void** Main()<br>15. {<br>16. Base obj = **new** Derived();<br>17. obj.show();}} |

*Figure 47. Late Binding in C#.*

| Code Listing 51: Late Binding in Java | Code Listing 52: Late Binding in Python |
|---|---|
| 1. **class** Base{<br>2. **public void** show(){}<br>3. }**class** Derived **extends**<br>4. Base {<br>5. **public void** show () {<br>6. **System.out.print**("child<br>7. ");<br>8. }}**public class** Main {<br>9. **public static void**<br>10. main(**String**[] args) {<br>11. Base objBase = **new**<br>12. Derived();<br>13. objBase.show();}} | 1. **class** Base:<br>2. **def** show(**self**):<br>3. **print**("Base")<br>4. **class**<br>5. Derived(Base):<br>6. **def** show(**self**):<br>7. **print**("child")<br>8. d =Derived()<br>9. d.show() |

*Figure 48.Late binding In Java and Python.*

**Enforce Override Keyword:** The common issue face in method overriding when someone override the method of base class but not tell compiler explicitly that method is being overridden then compiler assumed it is class method this can lead to runtime error [27]. The other common mistake a programmer can is the change the return type or number of parameters in overriding the base class method. To prevent these accidental errors languages, provide the override keyword. The main function of override keyword is to prevent unexpected function overloading. It is always a good practice to find errors at compile rather than run time [27].

C++ 11 supports the override keyword before that is it not available in C++ for method to be overridden it is compulsory that method is declared as virtual otherwise the compiler do the static binding Figure 49 illustrate the overriding without override keyword which is error prone and showing the importance of override keyword by raising run time exception to mismatch method signature [18]. Java and C# also support the override keyword but it is not compulsory in Java as well but highly recommend to write because then compiler will generate a compile time error if the method signature of base class is not matched by the signature provided by the subclass. In C# we can only override virtual and abstract method [27] [17].

C# provide two ways to overcome the method override issue one is to enforce the override keyword and other is the usage of new keyword [27] [17]. New keyword is feasible when programmer don't have the access of base class functions and are unable to override the base class methods and want to call the subclass implementation. C# also provide the override keyword and it is compulsory to write override keyword without the class do not override method Figure 50 shows the difference between the new and override keyword in C# [17].

**Code Listing 53**: Overriding in C++

| Without Override keyword | With override keyword |
|---|---|
| 1. **class** Base{<br>2. **public** :<br>3. **virtual void**<br>4. show(**int**<br>5. x){}<br>6. };<br>7. **class** Derived :<br>8. **public** Base {<br>9. **public** :<br>10. **void** show(**string**<br>11. x){}};<br>12. **int** main()<br>13. {<br>14. Derived obj;<br>15. obj.show("msg");<br>16. **return** 0;<br>17. } | 1. **class** Base{<br>2. **public** :<br>3. **virtual void** show(**int**<br>4. x){}<br>5. };<br>6. **class** Derived : **public**<br>7. Base {<br>8. **public** :<br>9. **void** show(**string** x)<br>10. **override** {}};<br>11. **int** main()<br>12. {<br>13. Derived obj;<br>14. obj.show("msg");<br>15. **return** 0;<br>16. } |

*Figure 49.Overriding in C++.*

**Code Listing 54**: Overriding in C#

| With new keyword | With override keyword |
|---|---|
| 1. **class** Derived : Base<br>2. {<br>3. **public new void**<br>4. show()<br>5. {<br>6. **Console**.**WriteLine**(<br>7. "child");<br>8. }<br>9. } | 1. **class** Base<br>2. {<br>3. **public virtual void**<br>4. show()<br>5. {<br>6. **Console**.**WriteLine**("B<br>7. as<br>8. e");}}<br>9. **class** Derived : Base<br>10. {<br>11. **public override void**<br>12. show()<br>13. {<br>14. **Console**.**WriteLine**("c<br>15. hild");}} |

*Figure 50.Difference of new and override keyword in C#.*

**Prevent the method from being overridden:** Sometime we don't want make whole class non inheritable. It is a good practice to make some methods non overridable which definition is not changed in the sub class. The non overridable methods resolve at the compile time. C++ older version do not support non overridable methods but a programmer can write private methods but after C++11 final keyword is added in the language Figure 51 (Code Listing 55 line 11) generate the compile time error because we are attempting to override the final method.

Java supports final and C# provide sealed keyword as well to make methods which cannot be override in the subclass Figure 51 (Code Listing 56) illustrate the method which is not overridable in subclasses of Derived class. Python don't support non overridable methods.

| Code Listing 55: Final Method in C++ | Code Listing 56: Sealed Method in C# |
|---|---|
| 1. **class** Base{<br>2. **public** :<br>3. **virtual void** show()<br>4. **final**{}<br>5. };<br>6. **class** Derived :**public**<br>7. Base<br>8. {<br>9. **public** :**void** show()<br>10. **override** {}};<br>11. **int** main()<br>12. {<br>13. Derived obj;<br>14. obj.show();<br>15. **return** 0;} | 1. **class** Base<br>2. {<br>3. **public virtual void**<br>4. show()<br>5. {**Console.WriteLine**<br>6. ("Base");}}<br>7. **class** Derived : Base {<br>8. **public sealed**<br>9. **override void**<br>10. show()<br>11. {<br>12. **Console.WriteLine**<br>13. ("child");}} |

*Figure 51. Non Overridable Methods in C++ and C#.*

**Operator Overloading:** Operator Overlading is the phenomenon in which we can change the standard working of an operator [18]. Operator Overloading is not much appreciated feature of programming world because it effects the overall simplicity of program and increase the complexity of program [15] [23]. Another potential harm of using operator overloading is if programmer overload the addition operator and perform subtraction it could lead to confusion.

Despite of all the issues related to operator overloading it could solve the many real-world problems more efficiently but with careful usage it is great feature that makes the object-oriented programming flexible [23]. We can overload an operator in C++ by using built in operator function as shown in Figure 52 (Code Listing 57) (line 4,7) [18] [19]. Operator overloading in C# is pretty much similar as in C++ the only difference is the use static keyword before operator function [17]. Operator overloading in C# as shown in Figure 53 (Code Listing 58) (line 4,10). Operator overloading in

Python has special object methods the name of these methods is start with double underscore. The calling of these methods is atomically called by the interpreter. We can change behavior of user defined datatypes by redefining these special methods Figure 54 (Code Listing 59) illustrate the redefining of special method for user defined types [22]. Java do not support operator overloading because of various reasons but poorly written code is one of the main issues. Java classes cannot reach the level of flexibility that classes of C++ provide because of the operator overloading feature. Operator overloading is the powerful feature provide by the languages if use with special care [23].

**Code Listing 57**: Operator Overloading in C++

1. **class** Number{
2. **public** : **int** x = 10;
3. **public** :
4.    **const** Number **operator** +(Number obj){
5.       Number temp;
6.       temp.x = **this**->x + obj.x;
7.       **return** temp;
8. }
9. };**int** main()
10. {
11. Number N1,N2;
12. Number N3 = N1 + N2;
13. **cout**<<N3.x;
14. **return** 0;
15. }

*Figure 52. Operator Overloading in C++.*

**Code Listing 58**: Operator Overloading in C#

1. **class** Number
2. {
3. **public int** x = 10;
4. **public static** Number **operator** +(Number n1 ,Number n2){
5. Number temp = **new** Number();
6. temp.x = n1.x + n2.x;
7. **return** temp;
8. }}
9. **class** HelloWorld{
10. **static void** Main()
11. {
12.    Number n1 = **new** Number();
13.    Number n2 = **new** Number();
14.    Number n3 = n1 + n2;
15. }}

*Figure 53. Operator Overloading in C#.*

**Code Listing 59**: Operator Overloading in Python

1. **class** Number:
2.    **def \_\_init\_\_**(**self**):
3.       self.num = 10
4.    **def \_\_add\_\_** (**self**, other):
5.       temp = Number()
6.       temp.num = **self**.num + other.num
7.       **return** temp
8. n1 = Number()
9. n2 =Number()
10. temp = n1 + n2

*Figure 54. Operator Overloading by using Special Method in Python.*

**Support interfaces:** In this section we will discuss the relation between classes and interfaces which is known as realization [27]. Interfaces is just like a contract the class who implement interfaces must implement all the method in the interface. Interface is just like a class who have collection of methods us to implement the behavior of class [18]. Interface is the important feature of languages which do not support multiple inheritance.

We can implement polymorphic interface by using abstract classes but there is a limitation because we can only have one base class at a time. But to solve real life problems we have multiple related classes in the systems another issue with the abstract classes the communication between non related classes is difficult. Interfaces solve the above problems with more efficient manner interfaces provide the polymorphic interface and we can implement multiple interfaces in one class from any package any namespace and any assembly implementation of multiple interfaces in Java as shown in Figure 55 (Code Listing 60).

Interfaces also formed the 'is a' relationship which implement the interface this is known as interface inheritance. We cannot instantiate interfaces but we can assign the reference of subclass which implements that interface but we are not allowed to call any class method. Java interfaces have fields which is implicitly final and static by default. Interfaces in Java also have special methods which is called default method. The default method has simple or common implementation inside the interface we can change these types of interfaces without having any sort of compile time and run time exception Figure 56 (Code Listing 61) (line 13,15) shows the interface reference calling the method and (line 3) shows the declaration of default method inside interface.

Java provide another type of interface which is called functional interface a functional interface has only one abstract method. Functional interface is used in anonymous classes and with lambda expression also called anonymous functions Figure 57 (Code Listing 62) line (12,16) shows the working of functional interfaces in Java. One subtle problem in Java interfaces can occur is a class implementing multiple interfaces and some of them have function with same name. Java handles if function have same signature in every interface then each interface have same definition but if a return type is different than Java throws a run time exception Figure 57 (Code Listing 63) (line 10,20) illustrate the name clashing of methods in interfaces [21][27].

C# interfaces support properties constants methods events indexers and properties. C# support method with concrete definition as well we can also declare properties inside the interface Figure 58 (Code Listing 64) illustrate the properties and default methods in C# interfaces. Like Java we can create the reference of interfaces in C# as well and call methods of interfaces. C# also provide the explicit interface implementation which is used to identify each interface method uniquely even multiple interfaces have method with same name which is an issue in Java interfaces Figure 59 (Code Listing 65) shows the explicit interface implementation in C#. An interface is also passed as the parameter to a function and interface can inherit other interfaces which is called base interfaces [17] [28]. Table 6 shows the evaluation of considered programming languages in terms of their support to polymorphism and realization

| Code Listing 60: Implementation of Multiple interfaces in Java |
|---|
| 1.  **interface** interface1{ <br> 2.  **public void** show(); <br> 3.  } <br> 4.  **interface** interface2 { <br> 5.  **void** hide(); <br> 6.  } <br> 7.  **class** Derived **implements** interface1,interface2 { <br> 8.  **public void** show() { <br> 9.    System.out.print("Show"); <br> 10. }   **public void** hide() { <br> 11.   System.out.print("Hide");} <br> 12. } |

*Figure 55. Multiple interfaces in Java.*

| Code Listing 61: Interface reference in Java |
|---|
| 1.  **interface** interface1{ <br> 2.  **public void** show(); <br> 3.  **default void** print(){ } <br> 4.  } <br> 5.  **class** Derived **implements** interface1 { <br> 6.  **public void** show() { <br> 7.  **System.out.print**("Show"); <br> 8.  } <br> 9.  **public void** classMethod() { } <br> 10. } <br> 11. **public class** Main { <br> 12. **public static void** main(**String**[] args) { <br> 13. interface1 obj = **new** Derived(); <br> 14. obj.show(); //ok <br> 15. obj.classMethod(); //compile time error <br> 16. }} |

*Figure 56. interface reference to subclass in Java.*

| Code Listing 62: Functional interfaces in Java | Code Listing 63: Method Name clashing in Java |
|---|---|
| 1. **interface**<br>2. interface1{<br>3.   **public void** show();<br>4.   }<br>5. **public class** Main {<br>6. **public static void**<br>7. main(**String**[] args) {<br>8. interface1 inter = () –<br>9. ><br>10. **System**.**out**.**print**("implementing show");<br>11. }} | 1. **interface**<br>2. interface1{<br>3.   **public void** show();<br>4.   }<br>5. **interface**<br>6. interface2 {<br>7.   **int** show();<br>8. }<br>9. **class** Derive<br>10. **implements**<br>11. interface1,interface2 {<br>12. **public void** show() {<br>13. **System**.**out**.**print**("Sho<br>14. w");}<br>15. **public void**<br>16. classMethod() {}} |

*Figure 57. functional interfaces and name clashing in interfaces in Java.*

**Code Listing 64**: Properties and Default Methods in C# interfaces

1. **interface** Myinterface
2. {
3. **String** Name { **get**; **set**; }
4. **public void** show();
5. **public void** print(){}
6. }
7. **class** MyClass: Myinterface{
8. **public String** Name { **get**;**set**;}
9. **public int** x = 10;
10. **public void** show()
11. {
12.   **Console**.**WriteLine**("method show");}
13. }

*Figure 58. implementation of interface properties and Default Methods in C#.*

**Code Listing 65**: Explicit interface implementation in C#

1. **interface** Myinterface
2. {**public void** show();}
3. **interface** MyString
4. {   **public void** show();}
5. **class** MyClass: Myinterface,MyString{
6. **void** Myinterface.show()
7. {**Console**.**WriteLine**("myinterface show");}
8. **void** MyString.show()
9. {
10. **Console**.**WriteLine**("myString show");}}
11. **class** HelloWorld
12. {
13.   **static void** Main(){
14. Myinterface myinter = **new** MyClass();
15. myinter.show();
16. MyString mystring = **new** MyClass();
17. mystring.show();}}

*Figure 59.Explicit interface implementation in C#.*

ENVIRONMENTAL FEATURES

In this section we briefly discuss the environmental features. The criteria to evaluate this environmental feature is same as the technical features. The study [14] define and evaluate some environmental features of programming languages but they are limited to functional and declarative languages. We will define such environmental features which related to object-oriented paradigm.

1) DEMAND IN THE INDUSTRY

In order to solve the real-world programming challenges the choice of language is very important. The majority of languages which is used in academia is because of their high demand in the industry [14]. In order to evaluate the industrial demand of the particular language we define the following parameters (I) total number of web searches for a particular language (II) total number of projects on GitHub (III) total number of job posts for a particular language.

The total number of web searches is managed by google trend. There are several web plat forms which can provide the popularity of a particular programming language. First is TIOBE index (https://www.tiobe.com/tiobe-index/) TIOBE index rank languages based on web searches google trends Table 7 shows the statistics of all three parameters. All of this statistical information helps us to find the current ranking of a particular language. Second, we use the (https://www.devjobsscanner.com/blog/top-8-most-demanded-languages-in-2022/) to find the job count of different languages. build statistics base on the job keyword e:g react jobs counted as JavaScript jobs. Lastly, we will collect the total number of active code repositories from GitHub. by using (https://githut.info/) stats of active repositories The GitHub ranks the language on starts and total pull request and total push request we are considering the push requests stats.

The index used in the evaluation provide us the quantitively data therefore we do not map these on our defined qualitative values. We will calculate these values in scoring function section when we calculate the overall score of the language.

*Table 6. Polymorphism and Realization*

| Language | Default dynamic binding | Enforce override keyword | Support prevents the method from being override | Support operator overloading | Support interfaces |
|---|---|---|---|---|---|
| C++ | Partially | Partially | Fully | Fully | No |
| Java | Fully | Mostly | Fully | No | Fully |
| Python | Mostly | No | No | Partially | No |
| C# | Mostly | Fully | Fully | Fully | Fully |

*Table 7. Demand in Industry.*

| Language | Total Web Searches | Active Repositories | Job Posts |
|---|---|---|---|
| C++ | 12.96% | 86505 | 212503 |
| Java | 13.23% | 222852 | 443508 |
| Python | 14.51% | 164852 | 515428 |
| C# | 8.21% | 56062 | 304892 |

## 2) CONTEMPORARY OBJECT ORIENTED FEATURES

In this section we will discuss the features which provide more flexibility and extensibility while writing programs in object-oriented paradigm. We will evaluate the contemporary features of a object oriented language by using the following parameters (I) support generics (II) support default exception handler (III) support explicit record classes.

**Generics :** The generic programming provide great feasibility when using the same algorithm for different types [19]. The use of templates and generics remove the redundant code from the program it can also convert the runt time errors to compile errors [4] [19]. In our considered languages every language support the concept of generics and templates but C++ and C# and Java generics provide type safety then dynamic languages the type mismatch errors can be detected on compile time rather run time which save cost to determine the type of parameter at run time Figure 60 (Code Listing 66) illustrate the generic list of vehicles which contain different types of vehicles. Java supports wild notation in order to specify method according to super types and sub types but C# do not support wild notation [4] [17] [19] [27].

Python is dynamic typed language it decides the type of variable at run time. but it can also support primitive data type. Dynamically type languages can be costly because of run time error checking and cannot be used to systems which are strictly typed [15] [30].

**Default Exception handler:** The unusual events and run time errors which caused the program to be terminated un expectedly are called exceptions. The part code which deals with exception is known as exception handler. Proper exception handling produces robust and fault tolerant systems [4] [27]. Java provide default exception handler which terminate the program properly by raising exceptions implicitly same is true for C# and Python. Java support another keyword finally which execute whether exception occur it can be used to release critical resource.

C++ exceptions are user defined which means they are raised explicitly and these exceptions are not related to handler every time. C++ throwing exceptions with primitive data types highly effect on the readability of program so its good practice to define classes for exceptions. C# exception handling is very similar to Java but C# do not support throw keyword. Python also support exception but instead of catch Python have except and one additional keyword else which is execute when no exception is found Python can also support raise which work like throw in C++ and Java [4] [15] [27].

**Records:** Sometimes we need classes to store data only and some simple operations that operate on data. The classes which is used to store only data is known as data classes or records [15]. Almost every high-level language provides such mechanism to create record. C++ provide value types structures [15]. Java and C# support record keyword to declare the records the records in Java are implicitly final. Record can implement interfaces and override their methods [31]. C# record are pretty much similar to Java except they can inherit other records C# also supports structures like C++ to create records. C# also support sealed and abstract records Figure 62 shows the records classes in Java [28].

Python uses data class decorator create data class. Python data classes defined implementation of methods like C3 and Java and also support initializers. Python data class also support inheritance of data classes to to other data classes but the initializer of base class cannot be called by the initializer of data class the calling of base Python provides a special method which called the base initializer after derived data class initializer is called Figure 63 shows the data classes in Python [32]. The evaluation of contemporary features is shown is
Table 8.

**Code Listing 66**: Generic ArrayList in Java

**Code Listing 66**: Generic List in Java

```
1.   class Truck {
2.   String name="big truck";
3.   }
4.   class Car{
5.   String name =" City";
6.   }public class Main {
7.   public static void main(String[] args) {
8.   Car objCar = new Car();
9.   List vehicles  = new ArrayList();
10.  vehicles .add(objCar);
11.  Truck t = new Truck();
12.  vehicles .add(t);
13.  }}
```
*Figure 60. Generic List in Java.*

**Code Listing 67**: Templates in C++
```
1.   template <class T>
2.   class Array{
3.   T arr[10];
4.   public :void add(T obj,int index)
5.   {
6.   arr[index] = obj;
7.   }
8.   };
9.   int main()
10.  {
11.  Array <int> mylist;
12.  mylist.add(1,0);
13.  mylist.add(2,1);
14.  }
```
*Figure 61. Template class in C++.*

**Code Listing 68**: Records in Java
```
1.   public record CarRecord(String name,int modal) {
2.   }
3.   public class Main {
4.   public static void main(String[] args) {
5.   CarRecord obj = new CarRecord("City",2001);
6.   CarRecord obj2 = new CarRecord("City",2001);
7.   System.out.print(obj.equals(obj2));//print true
8.   }}
```
*Figure 62. Record class in Java.*

**Code Listing 69**: Data Classes in Python
```
1.   from dataclasses import dataclass
2.   @dataclass
3.   class CarRecord:
4.   name:str
5.   model:int
6.   car1 = CarRecord()
7.   car1.name = "civic"
8.   car1.model = 2001
9.   car2 = CarRecord()
```
*Figure 63. Data class in Python.*

**Easy transferable**

Transfer of language is another important aspect of any programmer's life weather he is in industry or in academics. A good language should be easily transferable [4]. The syntax of Java C++ and C# are very much similar to each other so if someone has previous experience on one these languages can easily transfer himself on these languages. But if someone has experience on Python and any other dynamically type it would be difficult for him to learn the language quickly

In order to evaluate we define parameters paradigm shift, the sifting from imperative paradigm to object oriented have one unit cost and second is static type to dynamic language have one unit cost and strongly typed to weakly types language incur one unit cost. In order to find the minimum cost for each language we define the criteria which is shown in Table 9. Using the criteria define in Table 9 we construct the Table 10 and assign rating to each language cost.

**Foolproof integrated development environment**

The choice of ide is important in order to solve complicated and simple programming task. The right choice of ide makes programming task more efficient and less costly. In order to evaluate the integrated development environment, we use the following parameters (I) syntax checker (II) automatic code completion (III) support multiple languages (IV) support debugger.

Syntax checking at compile time can save time and cost. A programmer may write incorrect syntax for a particular expression and detection of these errors leads to reliable products. This will not only save time and cost as well. Availability of hints and auto completion is another important feature of ide. The auto completion and hints determine what type of arguments are function taking and definition of a particular function and complete the by using enter key **[4] [33]** .

Some ides are designed for one or two languages and some support multiple languages. DevC++ only supports the development of C/C++ programs but Visual Studio Code and Codelite and Netbeans supports development of multiple languages. Support of debugger is another important feature of ide. We can set the breakpoints in the code and determinethe value set by set execution. This feature really helps the new programmer to find errors it is highly recommend for the programmer to learn debugging skills **[4] [33] [34]**. The evaluation of foolproof IDE is shown in

Table 9. Criteria of calculating the cost of Language Transition.

| Fully | Mostly | Partially | No |
|---|---|---|---|
| Total cost <= 2N | 2N< Total cost <=2.5N | 2.5N < Total cost <=3N | 3N <Total cost |
| N= 4 Total cost <= 8 | 8< Total cost <=20 | 20 <Total cost <=24 | 24 <Total cost |

Table 10. Language Transition cost Calculation.

.

Table 8 Evaluation of Contemporary features

| Language | Support Generics | Support Default Exception Handler | Support Record Classes |
|---|---|---|---|
| C++ | Fully | No | Mostly |
| Java | Fully | Fully | Fully |
| Python | No | Partially | Partially |
| C# | Fully | Fully | Fully |

Consider a language L for which we need to compute the suitability score, Ls, based on its characteristics. As mentioned above, the proposed framework categorizes the evaluation criterion into two main categories, technical and environmental. However, while computing the score we have grouped all parameters in one block. We define the score of a language L against a parameter 'i' as ($Ls^i$). It might have been possible that some which results into variably different values for these parameters. Therefore, while mapping the qualitative measures onto the quantitative score, the resultant score of a parameter may become unbounded, as theoretically speaking, there may be any number of sub-parameters to evaluate a particular parameter. Furthermore, the parameters with wider range of possible scores may start overwhelming the other parameters. In order to restrict the score of each parameter in a closed interval,and

**Scoring function**

In this section we properly define the scoring of function for the evaluation of each object-oriented language. To find the quantified suitability score we use this scoring function to find the quantitative score for each considered language. Our proposed scoring function considers both technical and environmental features of the proposed framework, and assigns scores to a language based on its conformance to the criterion against each parameter. We map all four qualitative measurements for each considered parameter to a quantified score using values define in **Error! Reference source not found.**. These mappings is defined as Fully to 1, and No to 0 are very simple and intuitive, as 0 means no conformance, while 1 means full conformance to the criterion of a feature. In the same way, the other mappings are also supporting the criterion used for qualitative measurements as the mapping of Mostly to 0.70 reinforces the logic that majority of the features are being supported, and similarly, the mapping of Partially to 0.40 reflects that few of the requirements are justified and most of them are not supported by a language.

to avoid the aforementioned overwhelming affect, we normalize the score of such parameters by dividing the score of a parameter by maximum possible score for that parameter. As an example, the parameter ''FoolProof IDE'' is valuated on the basis of 4 sub-parameters, and for each parameter a language can have maximum score 1, thus the score obtained for this parameter is divided by 4. This results in restricting the score value for each parameter in [0,1] closed interval. In reality every user may have different priorities for each parameter. Therefore, we define a weight for each parameter which a user may assign to the parameter so as to prioritize it. As an example, one may be more interested in ''Fool Proof IDE'' as compared to the ''Easy Transferable'' of a language, in which case, the scoring function allows the user to assign a higher weight to one parameter and lower to the other. By default, each parameter

'i' carries weight 1, i.e. $w^{(i)} = 1$. The score for parameter 'i' is computed by multiplying the weight $w^{(i)}$ with the score of the parameter $Ls^{(i)}$, for the language L. Now, in order to compute the overall suitability score LS for a language L, we define a simple formula. This formula calculates the sum of score of every parameter and the final result is as follows

$$Ls = \sum_{i=0}^{n} w^{(i)} \cdot Ls^{(i)}$$

Where, 'n' is the total number of parameters in the language evaluation framework, which in our defined framework are 9. Ls gives us the suitability score for language L as an appropriate FPL. Hence, the above-mentioned scoring function, and discussion in the previous section help us computing the score for all languages, and the language with maximum suitability score turns out to be the most suitable FPL.

We have further processed the suitability score by dividing the obtained score by the sum of the weights of all parameters which helps restricting the overall suitability score in the [0,1] interval. This bounded or normalized score, with the default weight settings, implicitly reflects the overall percentage of conformance of a language to the proposed framework, i.e. 0.81 score reflects 81% conformance to the defined framework, similarly the difference of 0.02 should be treated as 2% less conformance. On the other hand, the benefit of using an unbounded score is that it reflects the differences in higher quantitative terms, but it fails to show the level of conformance to underlying proposed framework. We leave it to the user to choose any of the two score variants.

$$Ls' = Ls \Big/ \sum_{i=0}^{n} w^{(i)} \cdot Ls^{(i)}$$

Furthermore, in order to separately highlight the strength of a language from *technical* and *environmental* perspectives we have also computed *technical* and *environmental* scores in *unbounded* ($L_S^{TECH}$, $L_S^{ENV}$), and *normalized* ($L_S'^{TECH}$, $L_S'^{ENV}$) Versions, as shown in Table 12. Here, *'t'* is the number of *technical* parameters, and *'e'* is the number of *environmental* parameters in the framework, and $[t + e = n]$.

Table 11. Mapping of the qualitative measure onto the quantitative measure.

| Qualitative Measurement | Quantitative Score |
|---|---|
| Fully supported | 1 |
| Mostly supported | 0.70 |
| Partially supported | 0.40 |
| Not supported | 0 |

Table 9. Criteria of calculating the cost of Language Transition.

| Fully | Mostly | Partially | No |
|---|---|---|---|
| Total cost <= 2N | 2N< Total cost <=2.5N | 2.5N < Total cost <=3N | 3N <Total cost |
| N= 4 Total cost <= 8 | 8< Total cost <=20 | 20 <Total cost <=24 | 24 <Total cost |

Table 10. Language Transition cost Calculation.

| Language | C++ | Java | Python | C# | Total Cost | Rating |
|---|---|---|---|---|---|---|
| C++ |  | 1/0/0 | 1/1/1 | 1/0/0 | 5 | Fully |
| Java | 1/0/0 |  | 1/1/1 | 0/0/0 | 4 | Fully |
| Python | 1/1/1 | 1/1/1 |  | 1/1/1 | 9 | Mostly |
| C# | 1/0/0 | 0/0/0 | 1/1/1 |  | 4 | Fully |

*Table 12.Bounded and Unbounded Scores of a Language*

|  | Unbounded Scores | Bounded Scores |
|---|---|---|
| **Technical** | $Ls^{Tech} = \sum_{i=0}^{n} w^{(i)} . Ls^{(i)}$ | $L's_{TECH} = Ls / \sum_{i=0}^{n} w^{(i)} . Ls^{(i)}$ |
| **Environmental** | $Ls^{Env} = \sum_{i=0}^{n} w^{(i)} . Ls^{(i)}$ | $L's_{ENV} = Ls / \sum_{i=0}^{n} w^{(i)} . Ls^{(i)}$ |

**Scoring of Considered Languages**

In this section we formally compute the scores of our considered object-oriented languages by using the above scoring function. We calculate the scores of both technical and environmental features of the given language.

**Table 13** shows bounded and unbounded scores of technical features of each considered languages and environmental scores are shown in

Table **14**. The environmental parameter "Demand in *Industry"* are given a special consideration, as they are already presented in quantitative terms, so we have considered their quantitative values after bring the values to [0,1] interval, by dividing all values by the maximum for each sub-feature. This in turn, makes the score values of these features compatible with the rest of features.

Lastly, the scores of these features are combined while using the default weights in Table **15**. This table, in turn, shows the suitability score for each object-oriented language. From the calculation of suitability scores, it is clear that C# support more object-oriented principle than any other language. Technical features contain all the features and principle provided by object-oriented paradigm and non-technical and each principle of object-oriented programming is featuring especially technical features and in environmental features C# is one step below from Java. But industrial demands of C# are very low among the other languages.

As the default weight settings do not conform to the original popularity index of the languages, so there should be a different weighting criterion. However, it is very hard to come up with a generic and correct weighting criterion. For example, Python is very to full filling the construct of object-oriented programming but very high in industrial demand. So, a user can give high weightage to parameter "Demand in industry". We also calculate the scores on the base of higher industrial demand and from Table 16.Score **with higher weightage of Demand in Industry**

| Language | Ls | Ls' |
|---|---|---|
| Java | 5.61 | 0.51 |
| Python | 4.66 | 0.42 |
| C# | 4.23 | 0.38 |
| C++ | 3.81 | 0.34 |

we can see that Java and Python elevated and C# and C++ decrease in ranking. Hence, every user can look for an appropriate language based on her personal preferences.

However, based on the discussion in the previous section, it is clear that the user of this framework should have a reasonable understanding of the language theory to evaluate the language from *technical* perspective, and should have up-to-date information about tools, and statistics related to the language to evaluate *environmental* factors. But, the anticipated users of this framework are the personnel who are either course instructors, or curriculum designers, who in our opinion, possess sufficient background knowledge to use and customize such frameworks.

## VI. CONCLUSION

In this study we have proposed the detailed framework for the most widely used object-oriented languages on the basis of their implementation of pillars of object-oriented concepts. The research contains two categories technical and non-technical. We also define each feature and its sub feature and evaluate these sub feature. After evaluation we proposed the scoring function which is used to find the quantitative score of each object-oriented language. This is not only finding which language is best implementing the principle of object-oriented languages but it can also be used as guideline to design the new object-oriented language.

*Table 13. Scoring of Technical features of a Language (sorted by Ls)*

| Language | Abstract Datatype Encapsulation | Naming Encapsulation | Relationships among objects | Inheri |
|---|---|---|---|---|
| C# | 0.92 | 0.37 | 1 | 0.4 |
| C++ | 0.42 | 0.5 | 0.58 | 1 |
| Java | 0.42 | 0.35 | 0.85 | 0.4 |
| Python | 0.52 | 0.1 | 0.33 | 0.3 |

*Table 14. Score calculation of Environmental Features (sorted base on Ls)*

| Language | Demand in industry | Contemporary Features | Eas |
|---|---|---|---|
| Java | 0.92 | 1 | |
| C# | 0.46 | 1 | |
| Python | 0.91 | 0.46 | |
| C++ | 0.56 | 0.56 | |

*Table 15. Overall Score Calculation of each Language using default weight.*

| Language | Ls | Ls' |
|---|---|---|
| C# | 6.94 | 0.77 |
| Java | 6.59 | 0.73 |
| C++ | 5.75 | 0.63 |
| Python | 4.34 | 0.48 |

*Table 16. Score with higher weightage of Demand in Industry*

| Language | Ls | Ls' |
|---|---|---|
| Java | 5.61 | 0.51 |
| Python | 4.66 | 0.42 |
| C# | 4.23 | 0.38 |
| C++ | 3.81 | 0.34 |

30